\documentclass[12pt]{article}
\pdfoutput=1
\usepackage[english]{babel}
\usepackage[utf8]{inputenc}
\usepackage[T1]{fontenc}
\usepackage{amsmath}
\usepackage{amsthm}
\usepackage{amsfonts}
\usepackage{url}
\usepackage{authblk}
\usepackage{tensor}
\usepackage[bookmarks,colorlinks=false]{hyperref}
\usepackage{graphicx}
\usepackage{mathrsfs}
\usepackage{cite}
\usepackage{mathtools}
\usepackage{amssymb}

\newcounter{mnotecount}[section]

\newcommand{\beq}{\begin{eqnarray}}
\newcommand{\eeq}{\end{eqnarray}}
\newcommand{\ben}{\begin{eqnarray*}}
\newcommand{\een}{\end{eqnarray*}}

\newtheorem*{theorem*}{Theorem}

\DeclareMathOperator*{\wlim}{w-lim}

\begin{document}
\title{Einstein clusters as models of inhomogeneous spacetimes}
\author[1,3]{Sebastian J. Szybka}
\author[2,3]{Mieszko Rutkowski}
\affil[1]{Astronomical Observatory, Jagiellonian University}
\affil[2]{Institute of Physics, Jagiellonian University}
\affil[3]{Copernicus Center for Interdisciplinary Studies}
\date{}
\maketitle{}
\begin{abstract}
We study the effect of small-scale inhomogeneities for Einstein clusters. We construct a spherically symmetric static spacetime with small-scale radial inhomogeneities and propose the {\it Gedankenexperiment}. An hypothetical observer at the center constructs, using limited observational knowledge, a simplified homogeneous model of the configuration. An idealization introduces side effects. The inhomogeneous spacetime and the effective homogeneous spacetime are given by simple solutions to Einstein equations. They provide a basic toy-model for  studies of the effect of small-scale inhomogeneities in general relativity. We show that within our highly inhomogeneous model the effect of small-scale inhomogeneities remains small for a central observer. The homogeneous model fits very well to all hypothetical observations as long as their precision is not high enough to reveal a tension.
\end{abstract}

\section{Introduction}

The concept of idealization is one of the basic tools of modern physics. Macroscopic physical systems could be modelled only if unimportant details are neglected. Unfortunately, it is not always easy to decide which elements in the construction of the model are essential and which are not. It is believed that the decisive role is played by observational or experimental falsification. Again, this is not always straightforward. The most famous example is the model of our Universe. Its foundations have been proposed hundred years ago. This extremely simple model, which extrapolated by many orders of magnitude our faith in applicability of general relativity, turned out to be very successful. A hundred years later the model is alive and able to accommodate enormous flux of observational data provided by advances of modern technology. But, what some people see as a pure success for others is a failure. The model seems not to be free from strange coincidences and tensions. Moreover, $96\%$ of its energy content has not been previously known and is seen only via gravitational interactions. This apparent contradiction motivated broad studies of validity of a basic assumption of the model --- exact spatial homogeneity. In our article, we take on this topic and study an effect of small-scale inhomogeneities which is especially interesting in the light of the recent presumable tension between `local' and `early' universe measurements of the Hubble constant \cite{planck,riess,holicow,freedman}. These discrepancies renewed interest in  the role of inhomogeneities \cite{bolejko,adamek,macpherson,wojtak,kenworthy}.

The problem of small-scale inhomogeneities may be split into two topics: the effect of inhomogeneities on geodesics (light, gravitational waves, test bodies) which alters interpretation of our observations and the so-called backreaction effect which alters the structure of spacetime in a sense which will be explained below.

The backreaction problem is usually formulated as {\it a fitting problem} \cite{fitting}. In this approach, one asks how to fit an idealized solution to a realistic (`lumpy') spacetime. The aim of such approach is to find covariant procedure which uniquely assigns the best effective spacetime to a realistic one. However, it is more common in a down-to-earth scientific work to assume an effective model {\it a priori}. A physicist who wants to describe the complicated system usually neglects `details' and proposes a simplified model. This model is later being tested in experiments or against observations. In cosmology, spatial isotropy and homogeneity of the universe was a natural first guess. These assumptions led to our standard cosmological model $\Lambda CDM$. This model, with free parameters estimated by astronomers, constitutes the `effective spacetime.' Therefore, instead of looking for a fitting procedure one may formulate backreaction problem in an alternative way and ask what kind of errors has been introduced by idealization. 

In this alternative approach, the effective spacetime is known from the beginning. An idealised geometry does not fit to the matter content exactly and a discrepancy between the left hand side (geometry) and the right hand side of Einstein equations (the energy-matter content) arises. If one assumes that Einstein equations hold, then additional or missing terms are incorrectly interpreted as a contribution to the energy-momentum tensor. These artificial terms are known as a backreaction tensor. Since the $\Lambda CDM$ energy-momentum tensor is dominated by dark matter and dark energy --- the forms of energy and matter detected so far only through their gravitational interactions, then the backreaction effect has a potential to clarify our understanding of the Universe.

The results presented in the article by Stephen Green and Robert Wald \cite{greenwald} suggest that that this potential has not been realised in nature: within the formalism used there (the so-called Green-Wald framework) and under appropriate mathematical conditions the backreaction tensor is traceless, thus it may mimic radiation, but it cannot mimic cosmological constant nor cold dark matter. In the context of the $\Lambda CDM$ model this implies that backreaction effect introduces a minor correction  and it is definitely not the `order of magnitude effect' (which is needed to explain cosmological observations without cosmological constant or other forms of dark energy).

The Green-Wald framework \cite{greenwald} is a generalization to non-vacuum spacetimes and matter inhomogeneities of the Burnett's approach \cite{burnett}. (Burnett put on a rigorous mathematical footing the Isaacson's shortwave approximation \cite{isaacson1,isaacson2}.) One may, at least formally, find relevant examples of spacetimes with small--scale inhomogeneities, such that the Green-Wald formalism cannot be directly applied to them, e.g.\ a vacuum cosmological model with all the mass concentrated in a statistically homogeneously distributed black holes \cite{bhl}. Moreover, even if the backreaction vanishes, the effect of small-scale inhomogeneities may still alter interpretation of observations, as will be illustrated by our example. 

Our approach is based on a class of solutions to Einstein equations called an {\it Einstein cluster}. This type of solutions have interesting properties that allow for novel studies of the effect of small-scale inhomogeneities in an alternative setting. Our research is restricted to a simple toy-model and, as such, it is only indirectly relevant for cosmology (for other studies based on exact solutions see also \cite{Biswas:2007gi,Bolejko:2011jc,swisscheese,ourwm,ers,Sikora:2016,Sikora:2018}; for the current review of cosmological backreaction see \cite{coleyellis}).

\begin{figure}[t!]
\begin{center}
\includegraphics[width=10cm,angle=0]{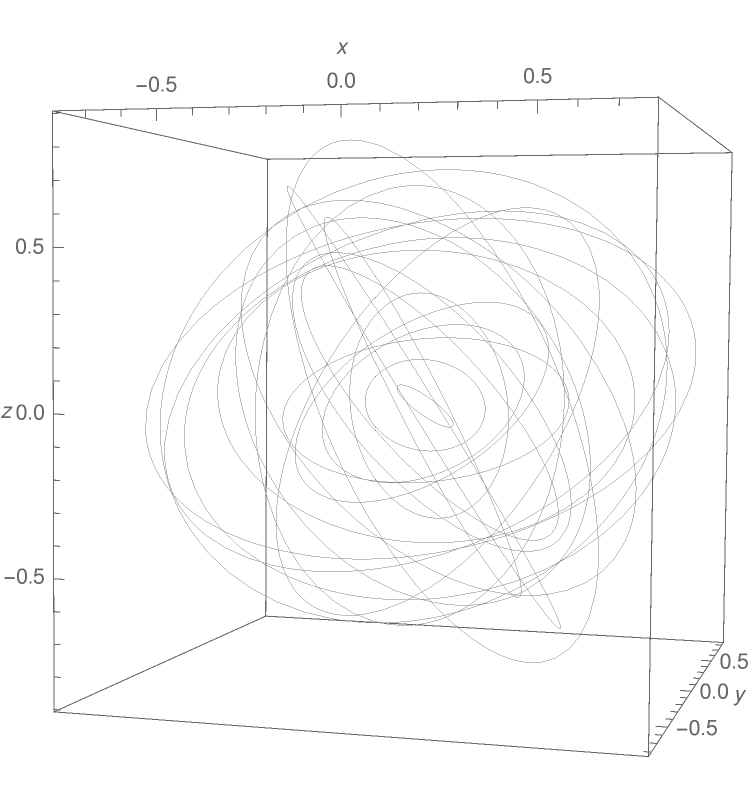}\label{fig:1}
\caption{The solution called `Einstein cluster' is a mean field approximation of a spacetime filled with a high number of massive particles moving in their own gravitational field on randomly inclined and directed circular orbits. (The Cartesian coordinates $x$, $y$, $z$ were scaled for simplicity.)}
\end{center}
\end{figure}

The {\it Einstein cluster} was discovered by Albert Einstein in 1939 \cite{einstein39}. It provides a mean field approximation of a cloud of massive particles moving in randomly inclined circular geodesics under the collective gravitational field of all the masses (see figure \ref{fig:1}). The spacetime is spherically symmetric and static. The radial pressure vanishes because the whole system is centrifugally supported.
Einstein clusters have been studied extensively in literature (see \cite{onec} and references therein). In the astrophysical context, they have been proposed as models of galactic dark matter haloes \cite{dmh}. 

The vanishing of radial pressure allows to construct static spacetimes with small-scale inhomogeneities without introducing unphysical equation of state. This property suits our purpose superbly. One of the interesting questions related to the effect of inhomogeneities is how large this effect might be. In the studies, as those presented in the article \cite{swisscheese}, large overdensities would lead in a short time to the formation of shell-crossing singularities, thus the magnitude of the effect is constrained by the construction of the model. This is an artifact of the simplified energy content which is pressureless (a dust). The solutions studied in our article are static and do not have this limitation. The effect of inhomogeneities is most interesting in the cosmological context where the structure formation may play a decisive role. However, it is not the aim of this research to clarify this aspect of backreaction. Our aims are more humble. We would like to verify the magnitude of the effect of small-scale inhomogeneities in the most simple static setting. This direction of studies seems to be unexplored so far. We point out that majority of authors by `backreaction' understand a dynamical effect. The effect studied here is of different nature: we study tensions in the homogeneous static model due to omission of the small-scale inhomogeneities.

Another novel property of the model studied is the fact in which the homogeneous spacetime is chosen. In the studies based on the swiss-cheese model, the inhomogeneous solutions are constructed out of the Einstein-de Sitter solution. Therefore, it is natural to compare the inhomogeneity effect against the Einstein-de Sitter background. However, this is a simplification. The observer living in a swiss-cheese universe would choose a different homogeneous model. Parameters of the homogeneous model would be based on observations in the swiss-cheese model interpreted under the homogeneity assumption. In this study, we elaborate on this topic in context of Einstein clusters.

The aim of the article is to conduct the {\it Gedankenexperiment}. We construct a solution to Einstein equations which contains small-scale inhomogeneities. Our choice of the density profile is inspired by the Green-Wald framework \cite{greenwald}: the metric is `close' to the homogeneous solution, but the second derivatives of the metric are large. We show that in our model the backreaction vanishes (in the sense of the Green-Wald framework \cite{greenwald}). Moreover, we present a heuristic analysis which implies that the backreaction vanishes in all possible models constructed within an Einstein cluster class. This motivate us to study the effect of inhomogeneities beyond the Green-Wald framework. We adopt a point of view of an astrophysicist who would like to model our inhomogeneous spacetime by available idealised exact solutions. We argue that astronomical observations interpreted within simplified model would lead to the misinterpretation of the energy content of the model. Our analysis is restricted to the particular class of solutions to Einstein equations, but it illustrates what the effect of small-scale inhomogeneities could be in principle.

\section{Setting}\label{sec:comparison}

Any spherically symmetric static spacetime could be written in the following form
\begin{equation}\label{eq:metric}
g=-e^\nu dt^2+e^\lambda dr^2+r^2(d\theta^2+\sin^2\theta d\varphi^2)\;,
\end{equation}
where $\nu$, $\lambda$ are functions of $r$ only. Two of the Killing fields could be immediately read out from the form of the metric: $\partial_t$, $\partial_\varphi$. 
For the centrifugally supported cloud of massive particles (the so-called Einstein cluster) the energy--momentum tensor non-vanishing components are \cite{einstein39}
\begin{equation}\nonumber
T^t_{\;t}=-\rho\;,\;\;T^\theta_{\;\theta}=T^\varphi_{\;\varphi}=p\;,
\end{equation}
where $\rho=\rho(r)$ is the energy density and $p=p(r)$ is a tangential pressure. The Einstein equations imply 
\begin{equation}\label{eq:lambdap}
\lambda=\ln (1+r \nu')\;,\;\;p=\frac{r\nu'}{4}\rho\;,
\end{equation}
and 
\begin{eqnarray}\label{eq:rho}
r \nu''+r(\nu')^2+2\nu'=8\pi r (1+r \nu')^2 \rho\;.
\end{eqnarray}
In order to find a particular solution one may set $\rho(r)$, solve \eqref{eq:rho} for $\nu(r)$ and calculate $\lambda(r)$ from \eqref{eq:lambdap}. The standard pseudopotential analysis reveals \cite{onec} that the radial stability conditions have forms
\begin{eqnarray}\label{eq:rsc}
&&0<r\nu'/2<1\;,\\
&&r\nu''-r(\nu')^2+3\nu'>0\;.\nonumber
\end{eqnarray}

The equation \eqref{eq:rho} if written in terms of an auxiliary function $\lambda=\ln\zeta$ [using \eqref{eq:lambdap}] reduces to the Bernoulli differential equation 
\begin{equation}\nonumber
\zeta'+P \zeta=Q \zeta^2\;,
\end{equation}
where $P=-1/r$, $Q=-1/r+8 \pi r \rho$. The substitution $\zeta\rightarrow 1/\mu$ leads to a linear equation of the form
\begin{equation}\nonumber
-\mu'+P \mu=Q\;,
\end{equation}
which has a solution 
\begin{equation}\label{eqmu}
\mu=1-\frac{8\pi}{r}\int \rho r^2 dr\;,
\end{equation}
where an integration constant is fixed by regularity at the center (it depends on the form of $\rho$).
Therefore, for a given density profile the solution to Einstein equations is given by
\begin{eqnarray}\label{eq:solnu}
\nu&=&\int \frac{dr}{r}\left(\frac{1}{\mu}-1\right)\;,\\\label{eq:sollambda}
\lambda&=&\ln{\frac{1}{\mu}}\;,
\end{eqnarray}
where $\mu$ may be calculated from \eqref{eqmu}. 

This solution may be matched to the Schwarzschild exterior. The active gravitational mass inside of the sphere with an area radius $r$ is given by \cite{onec} 
\begin{equation}\label{mass}
M(r)=4\pi \int_0^r \rho(\hat r) \hat{r}^2 d\hat r=\frac{r^2}{2}\frac{\nu'}{1+r \nu'}\;,
\end{equation}
and the total mass of all `particles' inside this sphere is\footnote{There is a misprint in formula $(41)$ in \cite{onec}. The minus sign in front of the integral is unnecessary.}
\begin{equation}\label{massT}
M_T(r)=\frac{1}{2\sqrt 2}\int_0^r \sqrt\mu \sqrt{3-1/\mu}\left(-\hat r\mu'/\mu+1/\mu-1\right)d\hat r\;.
\end{equation}
The difference between the total mass and the active gravitational mass tells us how much energy is needed to `disassemble' the cluster. The fractional binding energy may be calculated as follows
\begin{equation}\label{BE}
BE=\frac{M_T-M}{M_T}\;.
\end{equation}

\section{Small-scale inhomogeneities}

We assume that an energy density $\rho=\rho(r/l)$ is an oscillating function where $l$ is a constant. A small value  of $l$ corresponds to high frequency oscillations. Using equations \eqref{eqmu}, \eqref{eq:solnu}, \eqref{eq:sollambda} one may calculate metric functions $\nu(r)$, $\lambda(r)$ that correspond to $\rho(r/l)$. In this way, we construct a spacetime with small-scale inhomogeneities. 

It is not an aim of this paper to model any realistic astrophysical system, but in order to gain physical intuition one may pretend that our inhomogeneous spacetime describes the galactic halo. For simplicity, we choose $\rho(r/l)$ in such a way that it oscillates about a constant density $\rho_0$ for a class of stationary observers. Moreover, our system constitutes a finite configuration: at some radius $r=R$ it is matched to the Schwarzschild solution. 

We assume that a hypothetical astrophysicist living at the center of the system does not know $\rho$ precisely, but knows that $\rho$ is `approximately' constant and that the configuration is finite. Both facts would become basic assumptions of his idealised model: the Einstein cluster with a constant energy density and anisotropic pressure, from now on called the model $A$.

It follows from the Birkhoff theorem that vacuum spacetime outside a spherically symmetric configuration is given by the Schwarzschild metric. Thus, any effective spacetime must be also matched to the Schwarzschild solution. Imagine a hypothetical astrophysicist living in the center of this cluster who wants to determine its properties. Observations of trajectories of satellite stars and dwarf galaxies would allow to estimate gravitational mass of the system $M$. In addition, one may measure the blushift $z$ of the most distant stars (at the matching surface $r=R$). Alternatively, one may estimate the sum of all masses (stars, dark matter, \dots) that constitute the cluster $M_T$ or/and determine the angular diameter $d_A$ or the luminosity distance $d_L$ of the most distant objects.

To sum up, assumptions and hypothetical observational results which are made/known to our astrophysicist:
\begin{itemize}
\item the spacetime is static, 
\item the spacetime is spherically symmetric,
\item observer is at the center,
\item the matter is distributed uniformly on average (relatively to the preferred system of coordinates),
\item local effects are small (the observer and the sources are not in under/overdensities),
\item the configuration is finite (vacuum outside),
\item the state of art observations are not good enough to resolve individual inhomogeneities (their density profiles, etc.) --- the observer may detect only the cumulative effects,
\item one of three conditions holds:
\begin{itemize}
\item the gravitational mass $M$ of the configuration is known (based on observations of satellite dwarf galaxies and orbits of stars encircling the halo) and the blueshift $z$ of most distant stars in the halo is known (we assume that $z$ has been corrected for a perpendicular Doppler shift),
\item the gravitational mass $M$ is known and the observer measured the angular diameter distance $d_A$ of the most distant objects (or alternatively he/she measured the luminosity distance $d_L$),
\item the total mass $M_T$ of all constituents of the cluster is known and the observer knowns the angular diameter distance $d_A$ to the most distant objects.
\end{itemize}
\end{itemize}

It will be more instructive for a reader to start with the description of a constant density Einstein cluster (our effective and background spacetime --- our approach does not distinguish between these two concepts).

\section{Model $A$: constant density Einstein cluster}

A constant density profile $\rho(r)=\rho_A$ and the equation \eqref{eqmu} give (see also \cite{cdec})
\begin{equation}\nonumber
\mu_0=1-a_A r^2\;,
\end{equation}
where $a_A=8\pi\rho_A/3$ is a constant and where an additive constant was chosen to satisfy regularity at the center $r=0$. We have from \eqref{eq:solnu}, \eqref{eq:sollambda}

\begin{eqnarray}\nonumber
\nu_A&=&-\ln{\sqrt{1-a_A r^2}}+3\ln{\sqrt{1-a_A R_A^2}}\;,\\\nonumber
\lambda_A&=&-\ln\left({1-a_A r^2}\right)\;,
\end{eqnarray}
where without loss of generality we have chosen the additive constant \newline \mbox{$3\ln\sqrt{1-a_A R_A^2}$} in $\nu_0$ and where $R_A$ is a new constant $0<R_A<1/\sqrt{a_A}$. Finally, the metric reads
\begin{equation}\label{gA}
g_A=-\frac{\sqrt{1-a_A R_A^2}^3}{\sqrt{1-a_A r^2}}dt^2+\frac{1}{1-a_A r^2}dr^2+r^2d\Omega^2\;.
\end{equation}
The metric is regular and of the Lorentzian signature for $0\leq r < 1/\sqrt{a_A}$. The Ricci and Kretschmann scalars blow up at $r=1/\sqrt{a_A}$, so there is a curvature singularity. From now on we assume that $0\leq r \leq R_A < 1/\sqrt{a_A}$. For $r=R_A$ the spacetime is matched to the vacuum exterior Schwarzschild solution --- the active gravitational mass inside of the sphere with an area radius $r$ is given by \eqref{mass}. For a constant density profile $M(r)=a_A r^3/2$. The radial stability conditions \eqref{eq:rsc} reduce to
$0<3 a_A r^2<2$ and $a_A r^2<4/3$
which gives additional restriction on the matching hypersurface $r=R_A$.

\section{Inhomogeneous spacetime}

A toy-model studied in this paper is constructed as follows. We assume that $\rho(r)=2\rho_0 \cos^2(\pi r/l+\pi/4 \sigma)$, where $\rho_0$, $l$ and $\sigma=\pm 1$ are constant. The parameter $\rho_0$ is an average density as measured by stationary observers in our coordinate system and, at the same time, a local energy density at the center of the configuration. We introduce auxiliary constant $a$ such that $\rho_0=\frac{3a}{8\pi}$. The frequency of density oscillations is fixed by $l$ (the small value of this parameter $l\ll 1$ corresponds to high-frequency density oscillations). Note that an amplitude of these oscillations does not depend on $l$ and remains large in the limit $l\rightarrow 0$. For small $l$ our model represents highly inhomogeneous system! In order to interpreted $\rho_0$ as an average density it is necessary to assume that 
\begin{equation}\label{nl}
M(R)=\frac{4}{3}\pi R^3 \rho_0\;,
\end{equation}
which will be valid only for a discrete set of values of $R$.
The different choices of a phase $\pi/4\sigma$ (where $\sigma=\pm 1$) correspond to different matter configurations. The energy density at the center (at position of the observer) corresponds to its average value $\rho_0$.

We split $\mu$ into two parts: one which does not depend on $l$ and the second one which is $O(l)$: $\mu=\mu_0+\mu_l$. Using \eqref{eqmu} we find 
\begin{eqnarray}\nonumber
\mu_0&=&1-a r^2\;,\\\label{mul}
\mu_l&=&\frac{3al\sigma}{4 \pi^3}\left[-\frac{l^2}{r}+2\pi l\sin(\frac{2\pi r}{l})+(\frac{l^2}{r}-2\pi^2 r)\cos(\frac{2\pi r}{l}) \right]\;,
\end{eqnarray}
where an additive constant was chosen to satisfy regularity at the center. We have $\nu=\nu_0+\nu_l$, where from \eqref{eq:solnu}, \eqref{eq:sollambda}
\begin{eqnarray}\label{nu0}
\nu_0&=&-\ln{\sqrt{1-a r^2}}+3\ln{\sqrt{1-a R^2}}\;,\\\label{nul}
\nu_l&=&-\int \frac{dr}{r}\frac{\mu_l}{\mu_0}\frac{1}{(\mu_0+\mu_l)}\;,\\\nonumber
\lambda&=&-\ln\left(\mu_0+\mu_l\right)\;,
\end{eqnarray}
where the integration constant in $\nu_l$ should be chosen to satisfy $\nu_l(R)=0$. We have $g_{tt}=-e^{\nu_0+\nu_l}$, $g_{rr}=1/(\mu_0+\mu_l)$ and
\begin{equation}\label{g}
g=-\frac{\sqrt{1-a R^2}^3}{\sqrt{1-a r^2}}e^{\nu_l}dt^2+\frac{1}{1-a r^2+\mu_l}dr^2+r^2d\Omega^2\;.
\end{equation}
One may show that the Ricci and Kretschmann scalars blow up at \mbox{$\mu=0$,} so there is a curvature singularity. From now on we assume that that \mbox{$0\leq r \leq R < 1/\sqrt{a}$.} The term $\mu_l$ which is proportional to $l$ can be made arbitrary small, so the metric is regular and of the Lorentzian signature in $\mathbb{R}\times (0,\mathbb{R})\times S^2$. For $r=R$ the spacetime is matched to the vacuum exterior Schwarzschild solution --- the active gravitational mass inside of the sphere with the area radius $r$ is given by \eqref{mass}.

The radial stability conditions \eqref{eq:rsc} have complicated form for this solution. We have verified that they hold for our inhomogeneous density profiles for $R\gtrsim 6M$. The second radial stability condition may be marginally violated for some shells in certain more compact configurations, but this is not essential for our considerations.

\subsection{Green-Wald framework}

Our model corresponds to a one-parameter family of solutions to Einstein equations (with $l$ being a free parameter). One may verify by inspection\footnote{We put forward a hypothesis that any exact periodic density profile within the Einstein cluster class may be generalized to a one-parameter family of solutions satifying the Green-Wald assumptions.} that our model satisfies all assumptions of the Green-Wald framework \cite{greenwald} with the background spacetime $g^{(0)}$ which corresponds to $g_A$ with $a_A\rightarrow a$, $R_A\rightarrow R$, $\rho_A\rightarrow \rho_0$.

We define $h(l)=g(l)-g^{(0)}$.  The non-zero components of $h_{\alpha\beta}$ for small $l$ are
\begin{equation}\nonumber
h_{tt}\approx\-\frac{\nu_l}{\sqrt{1-a r^2}}\;,\;\; h_{rr}\approx-\frac{\mu_l}{(1-a r^2)^2}\;.
\end{equation}
It follows from the equations \eqref{mul}, \eqref{nul} that $\mu_l$, $\nu_l$ and the first derivatives of $\nu_l$ vanish in the high frequency limit $l\rightarrow 0$ (or $n\rightarrow\infty$). The derivative $\partial_r\mu_l$ is not pointwise convergent, but it remains bounded. We have $\lim_{l\rightarrow 0}h_{\alpha\beta}=0$ as expected. Although $\wlim_{l\rightarrow 0}(\nabla_\delta h_{\alpha\beta}\nabla_\gamma h_{\kappa\iota})$ does not vanish for $\delta=\alpha=\beta=\gamma=\kappa=\iota=r$, the backreaction tensor is zero ($\wlim$ denotes {\it a weak limit} as defined in \cite{greenwald} and a connection is associated with the spacetime $g^{(0)}$).

In summary, the one-parameter family of spacetimes \eqref{g} has a high frequency limit $\lim_{l\rightarrow 0} g=g_A$. It satisfies assumptions of the Green--Wald framework \cite{greenwald}. Although one component of $\nabla_\delta h_{\alpha\beta}$ is not pointwise convergent, the backreaction tensor vanishes. 

Vanishing of backreaction gives rise to another interesting question: Does there exist one-parameter families of solutions within Einstein cluster class [different choices of $\rho(r)$]  with non-trivial backreaction in the Green-Wald framework? We think that the answer to this question is {\it no}. We justify it as follows.

The possible source of backreaction is a nonlinear term $(\nu')^2$ in \eqref{eq:rho}. In order to be a source of the backreaction it would have to be non-zero in the high-frequency limit --- it should be at least $O(l^0)$. However, if $\nu'$ does not vanish for $l\rightarrow 0$, then it follows from \eqref{eq:lambdap} that $\lambda$ is not pointwise convergent which contradicts one of the Green-Wald assumptions about behavior of $h_{\alpha\beta}$ as $l\rightarrow 0$. Taking the high-frequency limit is a covariant procedure provided that the background (effective) spacetime is fixed. Therefore, all one parameter families of Einstein clusters to which the Green-Wald framework may be applied have vanishing backreaction.

\section{Gedankenexperiment}

Our inhomogeneous spacetime is defined by three parameters
\begin{itemize}
\item an average energy density $\rho_0$,
\item a size --- an area radius $R$,
\item a size of inhomogeneities $l$.
\end{itemize}
These parameters are fixed. The density $\rho_0$ must satisfy $\rho_0\leq 3/(8\pi R^2)$ to avoid curvature singularity.\footnote{In practice, we choose $M$ as a unit and calculate numerical value of $\rho_0$ from the equation \eqref{nl}.} The additional parameter $\sigma=\pm 1$ fixes phase of density perturbations. We will derive all our results for both values of $\sigma$ to grasp phase dependence. The effective model $A$ is defined by two parameters: $\rho_A$, $R_A$. 

Let $M$ be gravitational mass of the cluster, $M_T$ the total mass of its constituents and $z$, $d_A$, $d_L$, redshift, angular diameter distance, luminosity distance (respectively) of the most distant objects in the cluster. We assume that the observational data allow to determine one of the pairs: ($M$, $z$) or ($M$, $d_A$) or ($M$, $d_L$) or ($M_T$, $d_A$).

From \eqref{mass}, we have for the inhomogeneous spacetime
\begin{equation}\label{masses}
M(R)=\frac{4}{3}\pi R^3\rho_0-\frac{r}{2}\mu_l(R)\;,
\end{equation}
which together with the condition \eqref{nl} gives a transcendental equation for $R$, namely, $\mu_l(R)=0$. This equation does not depend on $\sigma$ and may be written as
\begin{equation}\nonumber
1-(1-\frac{x^2}{2})\cos(x)-x\sin(x)=0\;,
\end{equation}
where $x=2\pi R/l$. We solve this equation numerically. We skip three smallest roots and denote remaining subsequent roots with $x_k$, where $n\in \mathbb{N}$ and $x_k<x_{k+1}$. Using floor and ceiling functions we have $\lfloor \frac{x_k}{2\pi} \rfloor= \lceil k/2 \rceil$, thus $$2\pi<x_1<x_2<4\pi<x_3<x_4<6\pi<\dots\;.$$ Moreover, we define $R_k=x_k l/(2\pi)$.

Since the spacetime is spherically symmetric and static the blueshift $z$ for the inhomogeneous spacetime is given by
\begin{equation}\nonumber
1+z=\sqrt{\frac{g_{tt}(r=0)}{g_{tt}(r)}}=e^{\frac{\nu(0)-\nu(r)}{2}}\;,
\end{equation}
where $\nu(r)=\nu_0(r)+\nu_l(r)$ must be computed numerically from \eqref{nu0}, \eqref{nul}. The matching to the Schwarzschild solution implies $\nu(R_k)=\ln{(1-2M/R_k)}$.

For the effective spacetime $g_A$ [given by \eqref{gA}], the mass $M$ and the blueshift $z$ may be calculated as follows. Let $a_A=2M/R_A^3$, then at some $r=R_A$ the metric $g_A$ will match to the Schwarzschild solution. Since we have also $a_A=8\pi/3\rho_A$, then
\begin{equation}\label{massA}
M=\frac{4}{3}\pi R_A^3\rho_A\;.
\end{equation}
The blueshift is
\begin{equation}\label{zA}
1+z=\left(1-\frac{2Mr^2}{R_A^3}\right)^\frac{1}{4}\;.
\end{equation}
Finally, unknown parameters of the model $A$ (the effective spacetime), namely, $\rho_A$, $R_A$ in terms of a first pair $M$, $z$ of the `observational parameters' and parameters of the inhomogeneous spacetime $\rho_0$, $R_k$, $k$ are given by
\begin{eqnarray}\nonumber
\rho_A&=&\frac{3}{32\pi}\frac{\left((-z)(2+z)[(2+z)z+2]\right)^3}{M^2}=\frac{3^3}{2^9\pi^3}\frac{(1-\frac{e^{2\nu(0)}}{(1 - 8/3 \pi R_k^2 \rho_0)^2})^3}{\left(R_k^3\rho_0 \right)^2}\;,\\
R_A&=&\frac{2M}{-z(2+z)[(2+z)z+2]}=\frac{\frac 8 3 \pi R_k^3\rho_0}{1-\frac{e^{2\nu(0)}}{(1 - 8/3 \pi R_k^2 \rho_0)^2}}\;.\nonumber
\end{eqnarray}
It follows from the radial stability inequalities \eqref{eq:rsc} that the most compact stable/metastable configurations \cite{onec} in the homogeneous case correspond to $R_A=6M$, $R_A=3M$, respectively. The equation \eqref{zA} implies that the blueshifts for these configurations are given by $z=-1+(2/3)^{1/4}\simeq -0.096 $, $z=-1+1/3^{1/4}\simeq -0.240$, thus both systems are relativistic.

\begin{figure}[t!]
\begin{center}
\includegraphics[width=10cm,angle=0]{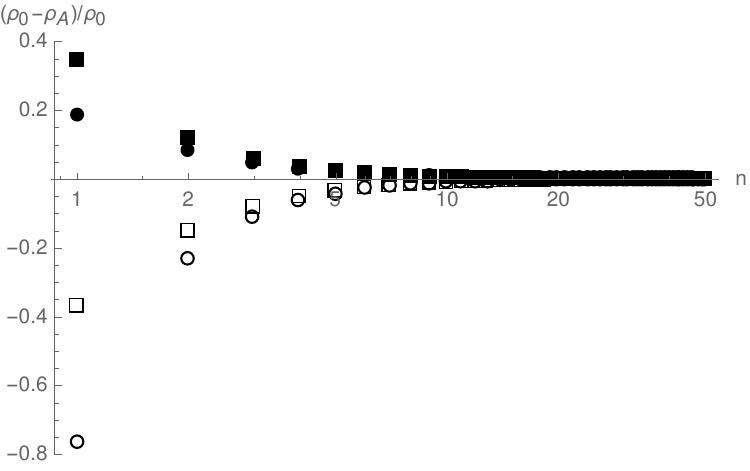}
\caption{The discrepancy between the average energy density $\rho_0$ in the inhomogeneous model and the estimated energy density in the homogeneous model $\rho_A$ for $R=4M$ as a function of a number of inhomogeneous regions. Circles and boxes correspond to different choices of a phase: $\sigma=1$, $\sigma=-1$, respectively. The filled symbols indicate that the outermost shell has higher than $\rho_0$ density (empty symbols indicate otherwise). The value of $\rho_A$ was estimated from the observed gravitational mass $M$ of the cluster and the redshift $z$ of outermost objects.}\label{fig:rho}
\end{center}
\end{figure}

Using observational data in the form of $M$ and $z$ within the homogeneous model the observer at the center will estimate $\rho_A$ to a different value than physical $\rho_0$. Therefore, other measurements of the energy density, i.e., from radiation that originates in decay of dark matter particles of the galactic halo (assuming that this will be known one day) will lead to a disagreement with $\rho_A$. We show for the configuration studied that the inhomogeneity effect vanishes in the limit of small inhomogeneities $l\rightarrow 0$. Let $n=\lfloor R/l \rfloor$ denote the number of inhomogeneous regions. The inhomogeneity effect as a function of $n$ is presented in figure \ref{fig:rho}. 
For $R=4M$ and $l=2\pi R/x_{100}$ (this implies $n=50$) we have at the boundary of the configuration $z\simeq-0.15908$ ($r=R$ corresponds to an overdensity), $z\simeq-0.15913$ ($r=R$ corresponds to an underdensity) for $\sigma=1$, $\sigma=-1$, respectively.

There are four points for each $n$ in figure \ref{fig:rho} (two choices of phase $\sigma=\pm 1$ and two values of $x_k$, $x_{k+1}$ such that $2\pi j<x_k<x_k+1<2\pi(j+1)$, where $j\in\mathbb{N}$). If $r=R$ corresponds to an underdensity/overdensity, then $\rho_A>\rho_0$, $\rho_A<\rho_0$, respectively. This may be understood intuitively in terms of photons and gravitational potential well. However, a careful  inspection of figure \ref{fig:rho} reveals that situation is not symmetric: an observer would most likely overestimate the local energy density based on many such observations. This asymmetry seems to be an artifact of the model studied.

Now, one can imagine an alternative procedure to determine the effective density. It follows from the form of the metric \eqref{eq:metric} that the angular diameter distance for radial beams and the central observer is $d_A=r$. If the observer knows gravitational mass of the cluster and angular diameter distance to the most distant astronomical objects in the cluster, then $R_A$ may be found directly as $R_A=R$. Since we assumed that the equality \eqref{masses} holds, then $\rho_A=\rho_0$ and the effect of inhomogeneities would be absent. However, if instead of angular diameter distance the luminosity distance to the most distant objects is known and the homogeneous model formula \eqref{zA} is used to calculate the redshift and estimate $R_A$, then a mismatch between $\rho_A$ and $\rho_0$ arises. This discrepancy is of similar nature, but slightly smaller amplitude than the one presented in figure \ref{fig:rho}.

\begin{figure}[t!]
\begin{center}
\includegraphics[width=10cm,angle=0]{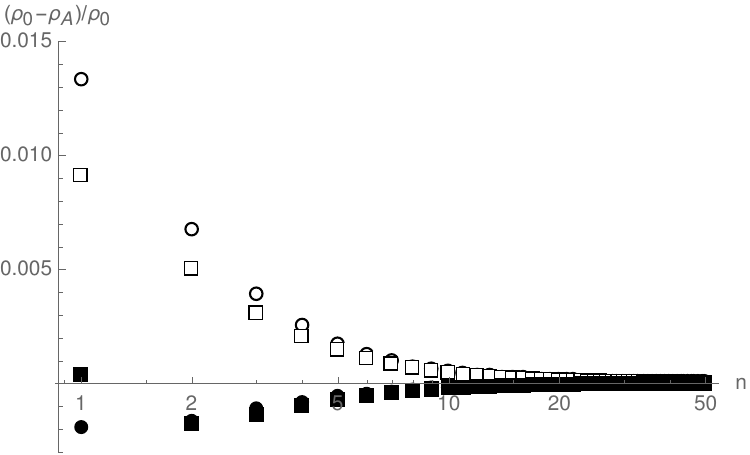}
	\caption{The discrepancy between the average energy density $\rho_0$ in the inhomogeneous model and the estimated energy density in the homogeneous model $\rho_A$ for $R=4M$ as a function of a number of inhomogeneous regions. Circles and boxes correspond to different choices of a phase: $\sigma=1$, $\sigma=-1$, respectively. The filled symbols indicate that the outermost shell has higher than $\rho_0$ density (empty symbols indicate otherwise). The value of $\rho_A$ was estimated from the observed total mass of cluster constituents $M_T$ and the angular diameter distance $d_A$ of outermost objects.}\label{fig:MTrho}
\end{center}
\end{figure}

Finally, one may consider different set of observations. Let us assume that instead of gravitational mass of the cluster $M$ an observer knows the sum of masses of all its constituents (e.g.\ stars, particles of dark matter,\dots) $M_T$. If the diameter distance to the most distant objects is known $d_A$, then the average density $\rho_A$ (under an assumption of homogeneity) may be calculated. The appropriate algebraic equation is too large to be usefully presented here. It may be derived as follows. The equation \eqref{massT} gives $M_T(r=d_A)$ in terms of $d_A$, $\rho_A$, $M$. The gravitational mass is unknown, but it may be written in terms of $d_A$ and $\rho_A$ under the homogeneity assumption [with the help of the formula \eqref{massA}]. The resulting algebraic equation may be solved numerically for $\rho_A$ in terms of $M_T$ and $d_A$. The results are presented in figure \ref{fig:MTrho}. Since the redshift $z$ was not involved in our calculations it may be expected that an outermost (near $R=d_A$)  underdensity/overdensity will not play decisive role. This did not turn out to be true. A heavy shell with large area radius enlarges the volume, so such configurations will have lower density than configurations for which the same amount of particles is contained in a lower volume $\rho_0<\rho_A$. The effect of outermost underdensities/overdensities is opposite to what has been observed in figure \ref{fig:rho} and the discrepancy is smaller. 

\begin{figure}[t!]
\begin{center}
\includegraphics[width=10cm,angle=0]{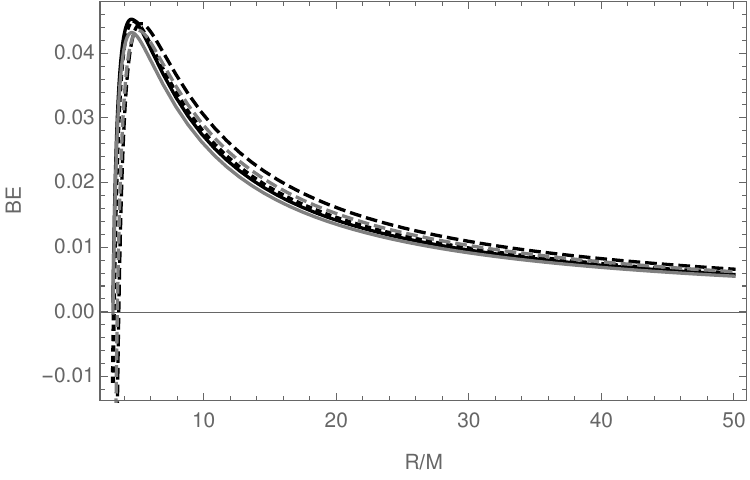}
\includegraphics[width=10cm,angle=0]{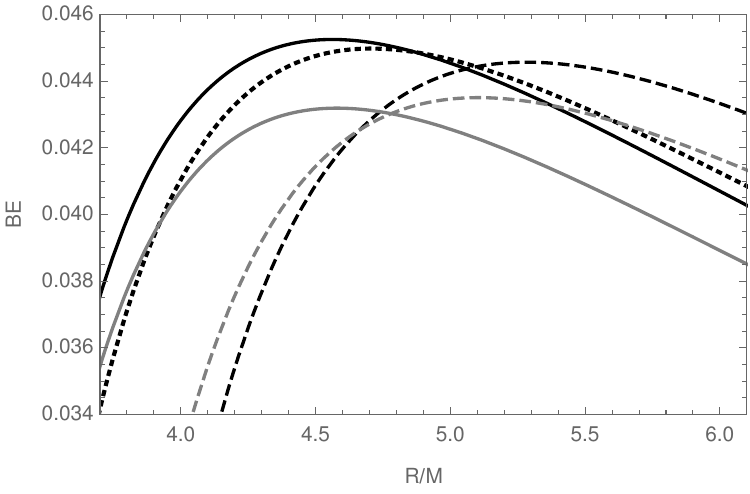}
	\caption{The fractional binding energy $BE$ as a function of compactness parameter $R/M$. The black/gray lines correspond to a phase $\sigma=1$, $\sigma=-1$, respectively. The solid/dashed lines correspond to the outermost inhomogeneity being overdesnity, underdensity, respectively.  All curves were plotted for $n=1$. The thick dotted line represent the homogeneous cluster (model A).}\label{fig:BE}
\end{center}
\end{figure}

In this paper, we investigate Einstein clusters with different density profiles. We approximate them by homogeneous Einstein cluster (model A) and study what kind of tensions are induced by such idealisation. These clusters may occupy different volumes and the outermost objects in them may have different redshifts. Another contribution to the effect of inhomogeneities is related to their fractional binding energies \eqref{BE}. We show in figure \ref{fig:BE} how fractional binding energies for different Einstein clusters of the same gravitational mass $M$ and the same coordinate radius $R$ depend on $R/M$. 

The fractional binding energy $BE$ for all clusters approaches asymptotically zero as $R/M\rightarrow +\infty$. It has a maximum between $R=4M$ and $R=6M$. The very compact configurations exhibit negative  binding energies. For small and large $R/M$, the fractional binding energy of the homogeneous cluster (dotted line) lies between fractional binding energies of clusters with outermost underdensities (dashed lines) and overdensities (solid lines). Their order flip near the maximum, but this cannot be directly seen in figure \ref{fig:MTrho} (it seems that the binding energy does not play a decisive role in the discrepancy of the densities presented in this figure).

\begin{figure}[t!]
\begin{center}
\includegraphics[width=10cm,angle=0]{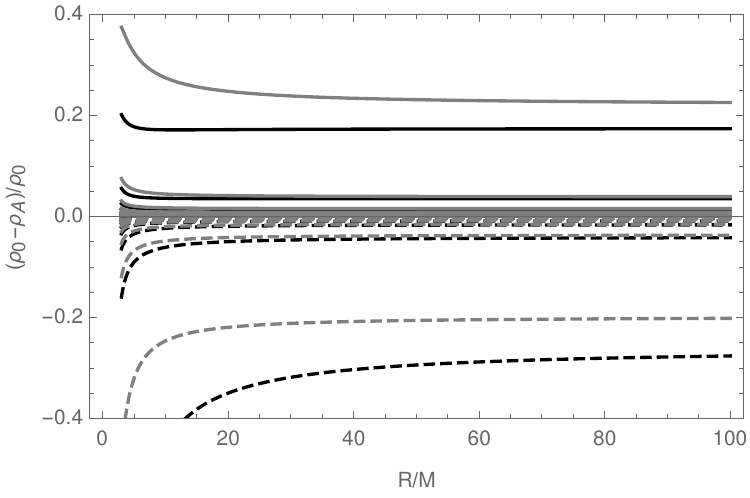}
\caption{The discrepancy between the average energy density $\rho_0$ in the inhomogeneous model and the estimated energy density in the homogeneous model $\rho_A$ as a function of compactness parameter $R/M$. The black/gray lines correspond to a phase $\sigma=1$, $\sigma=-1$, respectively. The solid/dashed lines correspond to the outermost inhomogeneity being overdesnity, underdensity, respectively. The curves were plotted for $n=1,3,\dots,21$. The amplitude of the effect of inhomogeneities decreases with growing number of inhomogeneities.}\label{fig:rhoRM}
\end{center}
\end{figure}

In our model, the effect of inhomogeneities for tens of inhomogeneous regions is not larger than a few percent. As the number of inhomogeneities grows, the effect vanishes (the limit $l\rightarrow 0$ or $n\rightarrow +\infty$) in accordance with the analysis within the Green-Wald framework (the high-frequency limit). The density contrast remains constant in this limit. What is interesting, the effect of inhomogeneities slightly decreases with the ratio $R/M$ (the area radius over the gravitational mass of the system) --- see figure \ref{fig:rhoRM}. Since there is no backreaction in the sense of the Green-Wald framework, the misinterpretation of the energy content is of trivial nature. It reduces to misinterpretation of the parameters of the model. A new form of the energy content cannot appear here because the effective spacetime belong to the same class of solutions to Einstein equations as the original one.

It was not an aim of our paper to model a realistic astrophysical system, but we find it instructive to calculate the effect of inhomogeneities for parameters corresponding to the dark matter halo of our Milky Way. We assume that in geometrized units the mass is $M=10^{12}M_\odot=1.477\times10^{15}m$ and the radius $R=4\times 10^5 ly=3.784\times10^{21}m$ which gives $R/M=2.563\times 10^6$. The Schwarzschild radius is one order smaller than stellar distances $2M=0.312 ly$. The energy density for the system compressed million times to the minimal configuration $R=3M$ would be $5.45\times 10^{-6} kg/m^3$ which still qualifies as a high vacuum for Earth standards. If the local clustering scale is assumed to be $l\approx 1kpc$ (the size of satellite dwarf galaxies), then $n\approx 40$. 
For these parameters the inhomogeneity effect (calculated from the gravitational mass $M$ and the redshift $z$) is small $|(\rho_0-\rho_A)/\rho_0|\approx 0.03\%$.

\section{Summary}

We have constructed the spherically symmetric static Einstein cluster with small-scale radial inhomogeneities and applied the Green-Wald framework to show that the effective energy-momentum tensor vanishes. Although the inhomogeneities did not generate artificial contribution to the energy-momentum tensor, they altered light propagation and observables. In order to quantify these effects, we extended our analysis beyond the Green-Wald framework. We have conducted the {\it Gedankenexperiment}: an observer at the center of this configuration modelled surrounding spacetime by an effective solution --- an homogeneous Einstein cluster. The parameters of this effective spacetime are based on straightforward astronomical `observations.' The idealization of the inhomogeneous spacetime resulted in the misinterpretation of the energy content. The effective energy density was different than the original average energy density. The sign of the effect depends on the configuration of matter studied and set of observables being used. The effect is not bigger than a few percent (assuming existence of more than a few inhomogeneous regions) and, as expected, it vanishes in the limit in which the size of inhomogeneous regions goes to zero, but the density contrast is kept constant. 

Our result does not generalize directly to the cosmological setting (which seems to be most interesting in the context of the inhomogeneity effect). Nevertheless, we have shown that in our model the homogeneous solution approximate quite well the inhomogeneous one, thus the effect of small-scale inhomogeneities is not an `order of magnitude effect', but it introduces a small correction. This conclusion is consistent with results obtained using different approaches \cite{Biswas:2007gi,swisscheese} (assuming  that local effects are negligible) and the current state-of-the-art of this topic (see the article \cite{coleyellis} and references therein).

Since the effect of inhomogeneities is not large, the different sets of not very precise observations would be initially consistent with the homogeneous model. The increased precision would reveal a tension between values of parameters based on different observations (under the homogeneity assumption). This tension will disappear only if the observational data would be reinterpreted within the inhomogeneous model.

\vspace{1.2cm}

\noindent{\sc Acknowledgments}

\vspace{0.2cm}

This publication was supported by the John Templeton Foundation Grant {\it Conceptual Problems in Unification Theories} (No.\ 60671). MR was supported by the Polish National Science Centre grant no.\ 2017/26/A/ST2/00530. Some calculations were carried out with the {\sc xAct} package \cite{xAct}. SJS thanks Krzysztof Głód and Szymon Sikora for a discussion.

\bibliographystyle{hunsrt}
\bibliography{report}

\end{document}